# Net Load Forecasting with Renewable Capacity Features: A Comparison of Direct and Indirect Neural Network Methods

Oluwafolajimi Samuel Bolusteve
Department of Electrical and Computer Engineering
University of Houston
Houston, TX, USA
osbolust@CougarNet.UH.EDU

Linhan Fang
Department of Electrical and Computer Engineering
University of Houston
Houston, TX, USA
lfang7@uh.edu

Xingpeng Li
Department of Electrical and Computer Engineering
University of Houston
Houston, TX, USA
xli82@uh.edu

***Abstract*—Renewable energy adoption has increased significantly in recent years. However, the growing penetration of renewable generation that depends on the weather introduces additional variability, making accurate net load forecasting increasingly challenging. In this study, long short-term memory (LSTM) and fully connected neural network (FCNN) models are employed to forecast net load one hour ahead using direct and indirect approaches. In the direct approach, net load is predicted directly, whereas in the indirect approach, total load, wind generation, and solar generation are predicted separately and then used to derive net load. Installed renewable generation capacity is also incorporated as an input feature to capture changes in renewable generation capability over time. A 24-hour input window and a chronological 90%/5%/5% training, validation, and test split are employed, with preprocessing fitted exclusively on the training data. On 1,754 common test timestamps, the direct FCNN slightly outperforms the indirect FCNN, whereas the indirect LSTM achieves the best overall performance, with a MAPE of 2.46%, an RMSPE of 3.59%, and an $R^2$ of 98.7%. These results demonstrate that component-wise forecasting can improve LSTM-based net load prediction when the characteristics of renewable generation and installed capacity information are represented appropriately.**

***Index Terms*—ERCOT, FCNN, LSTM, Machine learning, Neural network, Power prediction, Renewable capacity, Solar power, Texas power grid, Wind power.**

## I. Introduction

Global energy demand is expanding rapidly. The energy demand growth comes from increase in population, industrialization, recent increase in rate of data centers and so on [1]-[2]. For example, it is predicted the energy demand will increase by 3.3% in 2025 and a higher increase of 3.7% by 2026 [3]-[4]. As a result of this high energy demand and climate issues, the transition to renewable energy has gained growing attention [5]. The energy sources gaining popularity are solar and wind energy, which are being integrated into the grids to support the ever-increasing energy demand [6]-[8]. These renewable energy systems are meant to support and eventually replace energy sources that caused carbon emissions. The issue with solar and wind power is that the energy sources are difficult to predict since their energy sources are dependent on the weather. As a result, these energy sources are difficult to implement since they can result in energy outages and cause economic losses. Therefore, systems to predict energy production by wind and solar sources are required to prevent future outages [9]-[12].

In the literature, multiple forecasting approaches have been used, such as statistical models, machine learning models, and so on. For this research, machine learning models are used because those models predict better than statistical models like ARIMA [13]. Multiple machine learning models have been applied for prediction purposes, especially regarding energy demand. Commonly used approaches include Long short-term memory (LSTM), fully connected neural networks (FCNNs), gated recurrent units (GRUs) and support vector machines (SVMs) and others [14]-[17].

The FCNN is a feed-forward neural network architecture where each neuron from the previous layer is connected to every neuron from the current layer. The fully connected layers help with prediction because every input feature can influence other input features which is useful in load predictions [18]. LSTMs are enhanced recurrent architectures of RNNs designed to address the vanishing gradient of RNN. LSTM utilizes memory cells which can be used to predict to capture long-term patterns in sequential data. This is useful in load predictions because the notices of LSTM change over long periods which is useful in predictions. GRUs are simplified variants of LSTM which reduce computational cost since LSTM requires more computational power than most models. To reduce the computational cost, the forget and input gates are merged into one single gate. This method is more efficient while maintaining similar prediction as LSTM [19]-[20].

For data collection, multiple datasets were obtained from two main sources, Electric Reliability Council of Texas, Inc. (ERCOT) [21], and North American Land Data Assimilation System (NLDAS) [22]. Both are authoritative data sources that provide accurate and well-maintained datasets for energy research. Time-series data for system load, solar generation, and wind generation were obtained from ERCOT. Meteorological time-series data, including wind speed, irradiance, and temperature, were obtained from NLDAS.

Two main methods used for this model prediction are indirect and direct net load prediction. For the direct net load prediction, the model predicts only the net load, which will be compared to the indirect net load prediction. The indirect net load prediction predicts the solar energy produced, wind energy produced and the system load. The results are calculated using the net load calculation. Both results are compared using statistical methods.

The remainder of this paper is organized as follows. Section II introduces the methodology of the study along with a description of method used, graphical analysis of data and ML models used. Section III presents the results from machine prediction. Section IV concludes the paper.

## II. METHODOLOGY

This section consists of the following subsections: the data description, model parameters, training and validation, and the direct and indirect methods. This study focuses on FCNN and LSTM to provide a controlled comparison between direct and indirect net load forecasting strategies. FCNN is used as a feed-forward neural-network baseline, while LSTM is used as a recurrent architecture that captures temporal dependencies in hourly load and renewable generation profiles. Therefore, the goal of this work is not to provide an exhaustive model benchmark, but to evaluate how the forecasting formulation and renewable-capacity-aware features affect net load prediction performance under comparable neural network settings. Broader baselines, such as GRU, ARIMA, XGBoost, SVM, and Transformer-based models, will be investigated in future work.

### A. *Data Description*

The total load, solar, and wind power generation data from ERCOT for the years 2021 through 2024 are used to train and validate the machine learning models. The dataset features an hourly temporal resolution. In addition to the actual wind and solar power generation, it includes the total installed capacity, which represents the theoretical maximum power output of the operational turbines and solar panels.

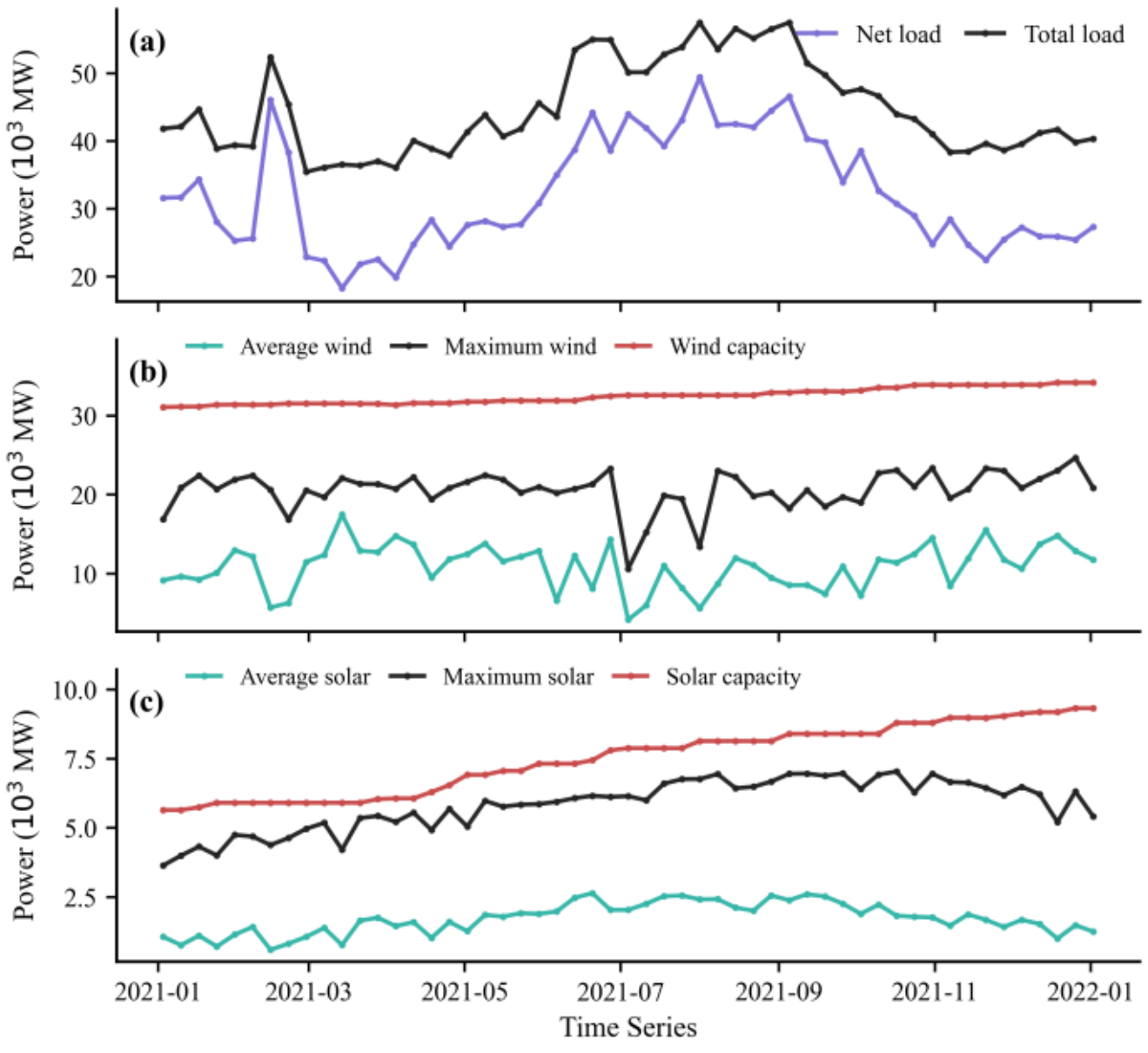

Fig. 1. Weekly aggregated profiles for 2021: (a) average net and total load; (b) average and maximum wind generation and installed wind capacity; and (c) average and maximum solar generation and installed solar capacity.

The meteorological variables include temperature, wind speed, and irradiance obtained from NLDAS. The NLDAS timestamps were converted from Coordinated Universal Time (UTC) to Central Standard Time (CST) to align them with the ERCOT timestamps.

As illustrated in Fig. 1 and Fig. 2, Fig. 1 presents weekly aggregated profiles for 2021, whereas Fig. 2 presents monthly aggregated profiles for 2021–2024. These figures illustrate short-term variability and long-term trends in system demand and renewable generation.

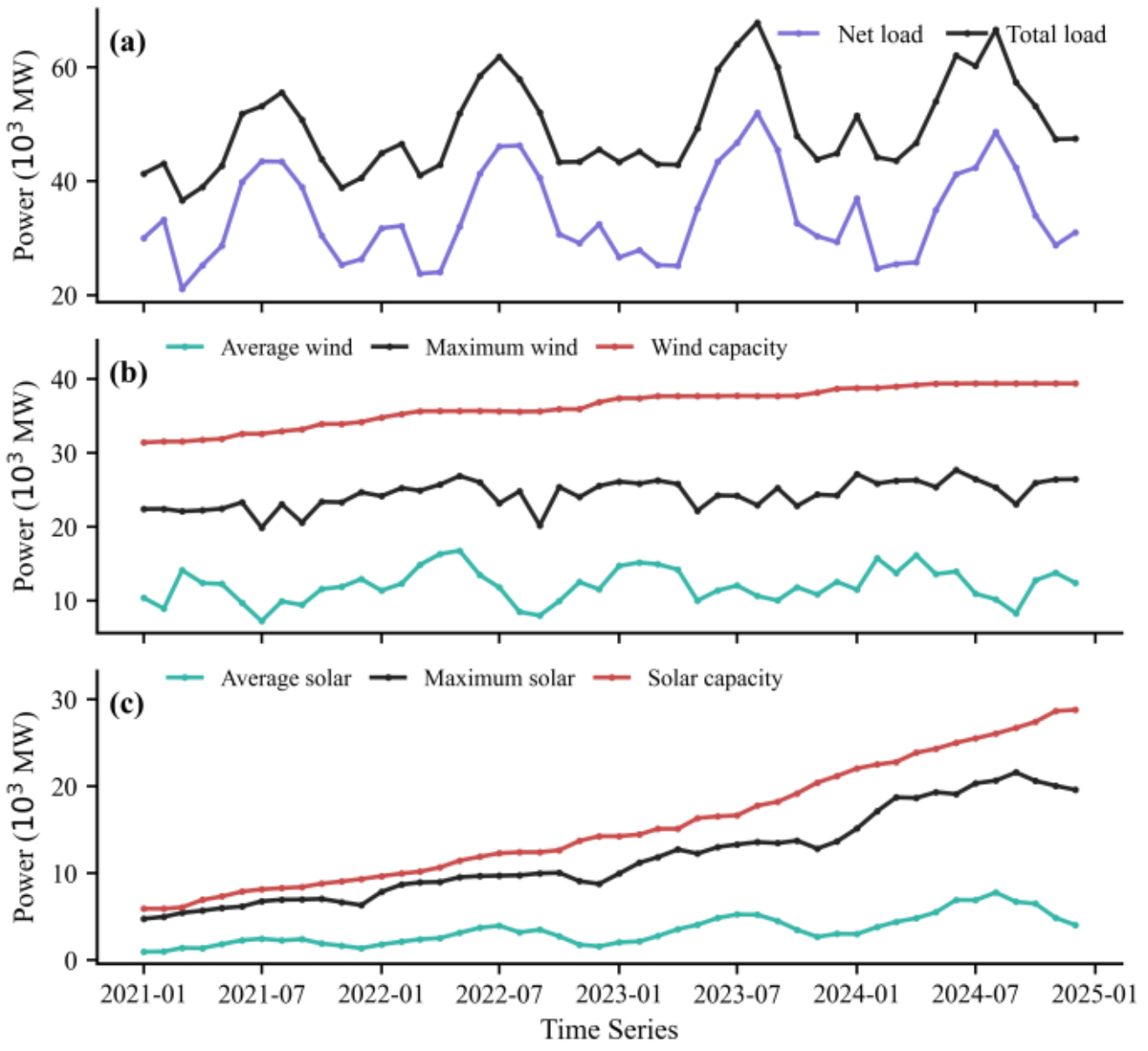

Fig. 2. Monthly aggregated profiles for 2021–2024: (a) average net and total load; (b) average and maximum wind generation and installed wind capacity; and (c) average and maximum solar generation and installed solar capacity.

The monthly aggregated data in Fig. 2 reveals a sustained year-over-year upward trajectory in total load. Furthermore, the widening gap between the total and net load visually underscores the continuously increasing penetration of renewable energy. The substantial and continuous expansion of installed solar capacity has driven a commensurate increase in both average and peak solar generation.

In contrast, installed wind capacity exhibits a more moderate, linear growth trajectory. Furthermore, the weekly resolution profiles depicted in Fig. 1 highlight the inherent high volatility of wind generation. Specifically, wind output frequently dips during the critical summer peak-demand periods, while maintaining higher generation levels throughout the spring, autumn, and winter.

### B. *Model Setting, Training and Validation*

This study employed both LSTM and FCNN models. Although the FCNN offers faster prediction speeds, the LSTM compensates for its longer computation time through its superior prediction accuracy. The internal structure of the LSTM is illustrated in Fig. 3.

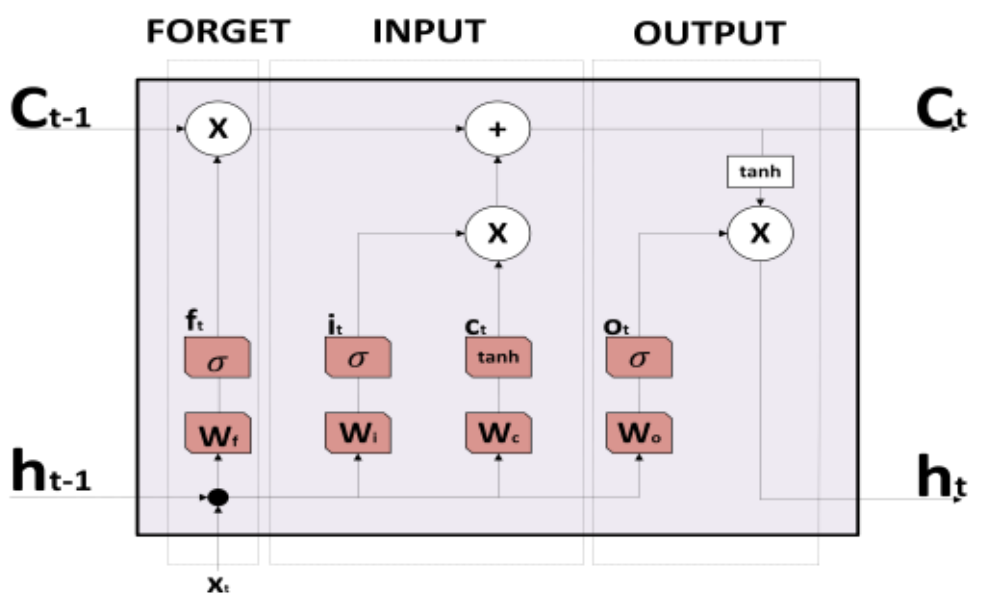

Fig. 3. Internal structure of an LSTM cell.

Both the LSTM and FCNN models were structured with two hidden layers and utilized a dropout rate ranging from 0.1 to 0.3, terminating in a final layer for prediction. The model employed Huber loss for robustness against outliers and an Adam optimizer with an initial learning rate of 0.0003. To monitor model performance and generalization, mean absolute error (MAE) and mean squared error (MSE) were tracked across both the training and validation sets, making this setup highly suitable for time series forecasting. Between these two models, some parameter settings were similar which are the window sizes, epochs, and training/validation/test split ratio.

The models employ a 24-hour historical look-back window to predict the power generation and load demand. This specific look-back period was designated to effectively capture the inherent diurnal periodicity and regular patterns of wind and solar power generation. Furthermore, as corroborated by the sensitivity analysis in Table I, a 24-h look-back window with a one-hour forecast horizon achieves the optimal balance between predictive accuracy and model efficiency.

Table I. Validation performance of the FCNN model across different prediction horizons

| Look-ahead Hour | MAPE (%) | RMSPE (%) | $R^2$ |
|---|---|---|---|
| **1** | **0.7479** | **0.957** | **0.998223** |
| 2 | 0.9462 | 1.24933 | 0.996884 |
| 4 | 1.1473 | 1.5067 | 0.99523 |

As the look-ahead horizon increases, the model's prediction accuracy and generalization capability decrease accordingly. The hourly samples were sorted by timestamp and split chronologically without random shuffling. After constructing the 24-hour input windows with one-hour-ahead targets, 35,065 valid samples were obtained and divided into training samples, validation samples, and testing samples based on a 90:5:5 ratio. This split was selected because the four-year hourly dataset contains both seasonal variations and long-term renewable capacity growth. Therefore, a larger training portion helps the models learn representative load and renewable generation patterns, while the validation and testing sets are reserved for model selection and out-of-sample evaluation.

To evaluate the performance of the machine learning models, three metrics were employed: the mean absolute percentage error (MAPE), Coefficient of Determination ($R^2$), and root mean squared percentage error (RMSPE). The prediction error is defined as the difference between the predicted and actual values. Specifically, MAPE quantifies the overall forecasting accuracy by averaging the absolute percentage deviations, mathematically expressed as follows:

$$\text{MAPE} = \frac{1}{n}\sum_{i=1}^{n}\left|\frac{y_i - \hat{y}_i}{y_i}\right| \times 100 \quad (1)$$

The coefficient of determination, $R^2$, measures the robustness of the model. The value ranges from 0 to 1, and the higher the value, the better the model prediction. The calculation for the $R^2$ score would be done by Sklearn library. RMSPE is a statistical measure that measures the average deviation between predicted values and actual values expressed in percentage.

$$\text{RMSPE} = \sqrt{\frac{1}{N}\sum_{i=1}^{N}\left(\frac{y_i - \hat{y}_i}{y_i}\right)^2} \quad (2)$$

## C. Direct and Indirect Methods

There were two approaches taken to predict the net load, as stated earlier. For the direct method, the model directly predicts the one-hour-ahead net load using the historical 24-hour input window, as shown in Fig. 4. The input features include wind speed, wind generation, wind capacity, total load, solar irradiance, solar generation, solar capacity, and temperature.

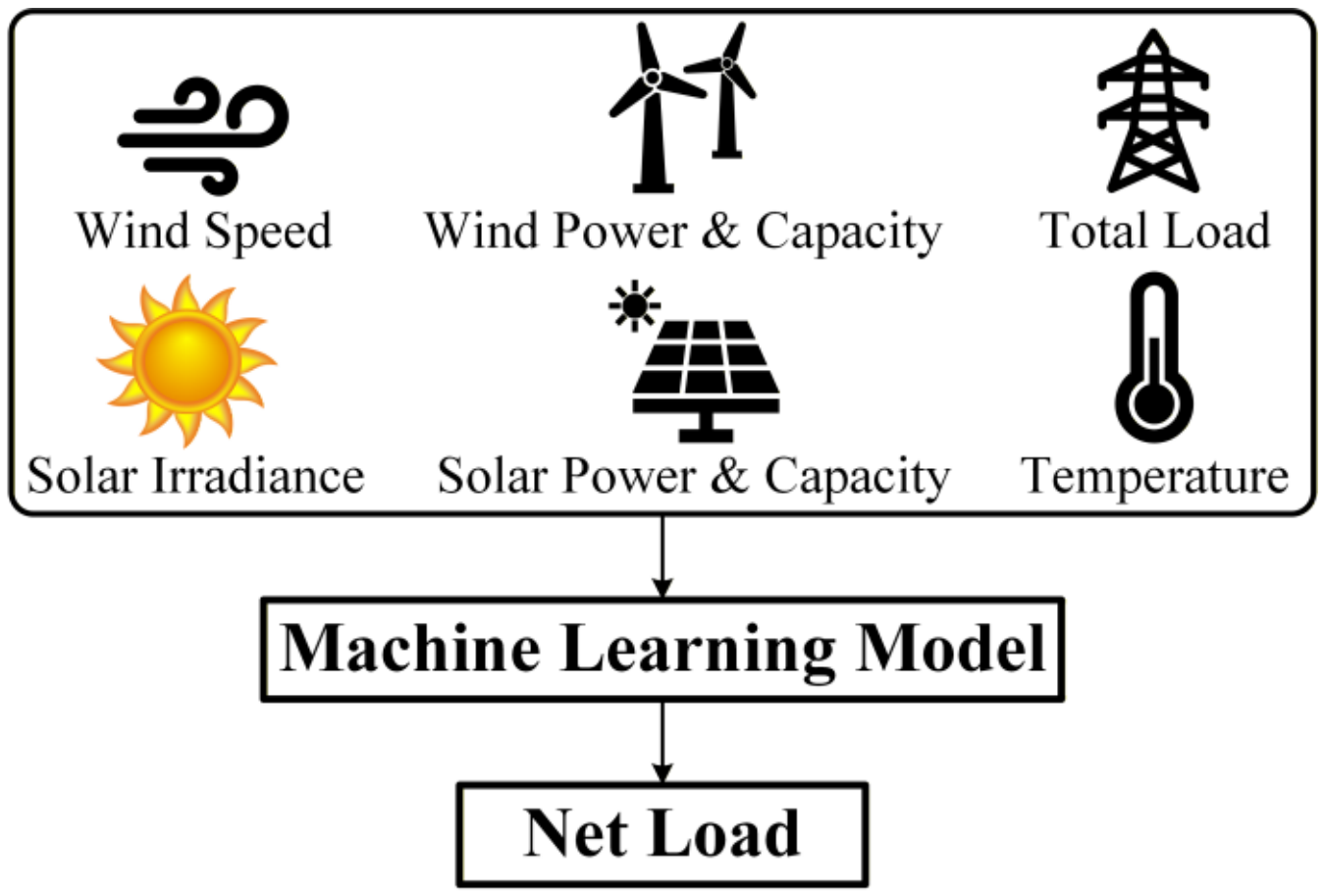

Fig. 4. Process for direct prediction of net load.

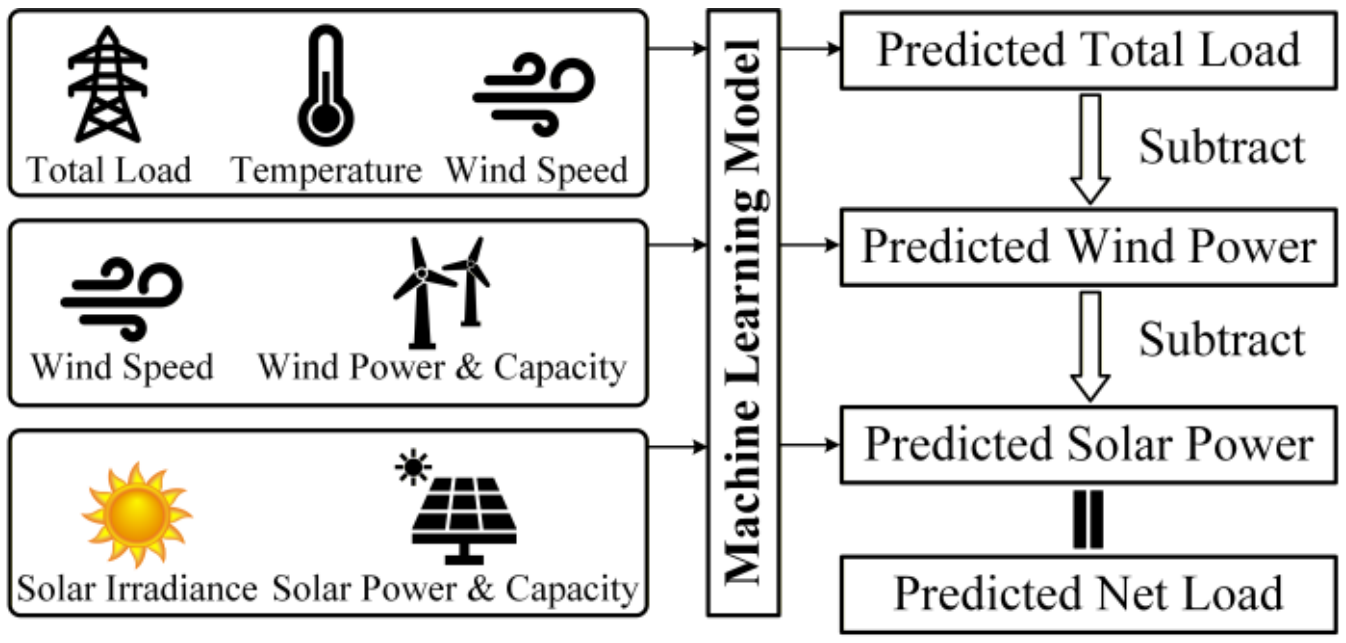

Fig. 5. Process for indirect prediction of net load.

For the indirect method, three separate models are trained to predict total load, wind generation, and solar generation, respectively, as shown in Fig. 5. The predicted net load is then obtained by subtracting the predicted wind and solar generation from the predicted total load. This formulation decomposes net load forecasting into physically interpretable components and allows each model to learn the target-specific relationship between weather variables, renewable capacity, and power output.

## III. Results

## A. Results of FCNN Model

The direct FCNN predicts net load as a single target, whereas the indirect FCNN separately predicts total load, solar generation, and wind generation before reconstructing net load. Their performance is compared using the actual and predicted profiles, MAPE, RMSPE, $R^2$, and signed percentage errors. Table II shows comparable performance between direct FCNN and indirect FCNN models, although the direct method produced modestly lower MAPE and RMSPE values and a higher $R^2$. The direct and indirect FCNN methods explained approximately 92.3% and 91.1% of the variance in net load, respectively.

Table II. FCNN Performance of both Direct and Indirect methods

| Performance Measure | Direct Method | Indirect Method |
|---|---|---|
| MAPE (%) | 7.14% | 7.40% |
| RMSPE (%) | 9.49% | 9.80% |
| $R^2$ | 92.3% | 91.1% |

Fig. 6 shows the actual and predicted net load profiles of the direct FCNN method over the first 300 test samples. The direct FCNN achieved a MAPE of 7.14%, an RMSPE of 9.49%, and an $R^2$ of 92.3%. Fig. 7 presents the corresponding results for the indirect FCNN method, which achieved a MAPE of 7.40%, an RMSPE of 9.80%, and an $R^2$ of 91.1% as shown in Table II.

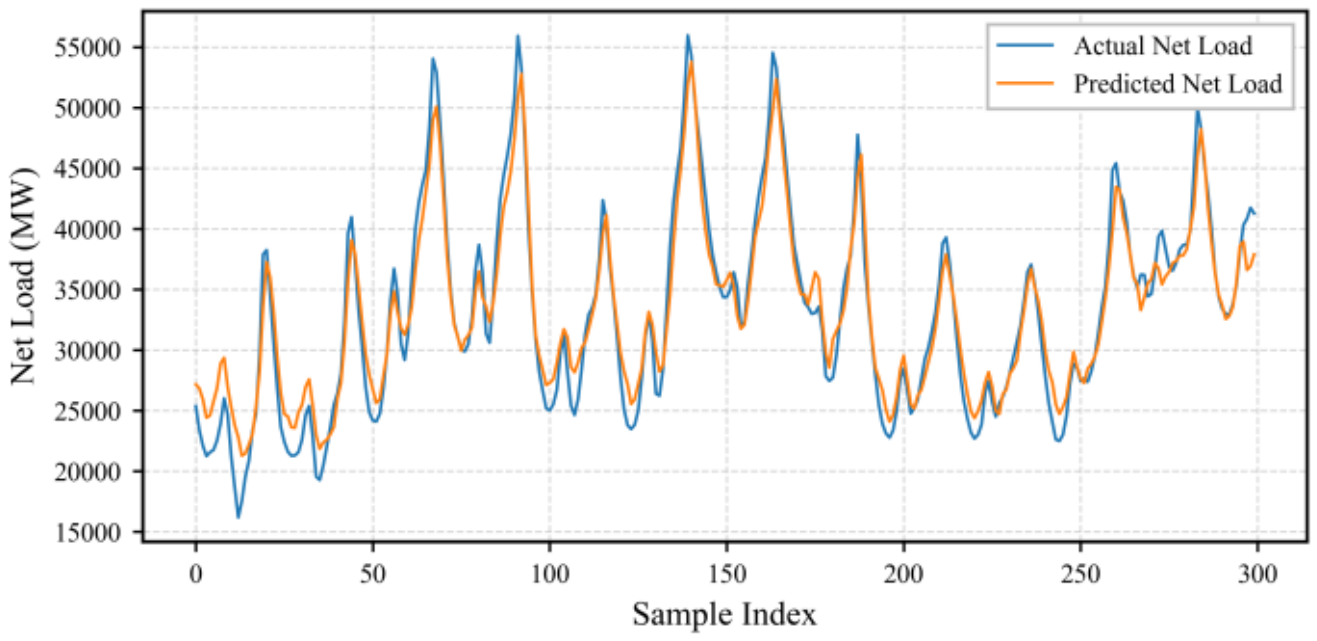

Fig. 6. Comparison of actual and predicted net load for direct FCNN method.

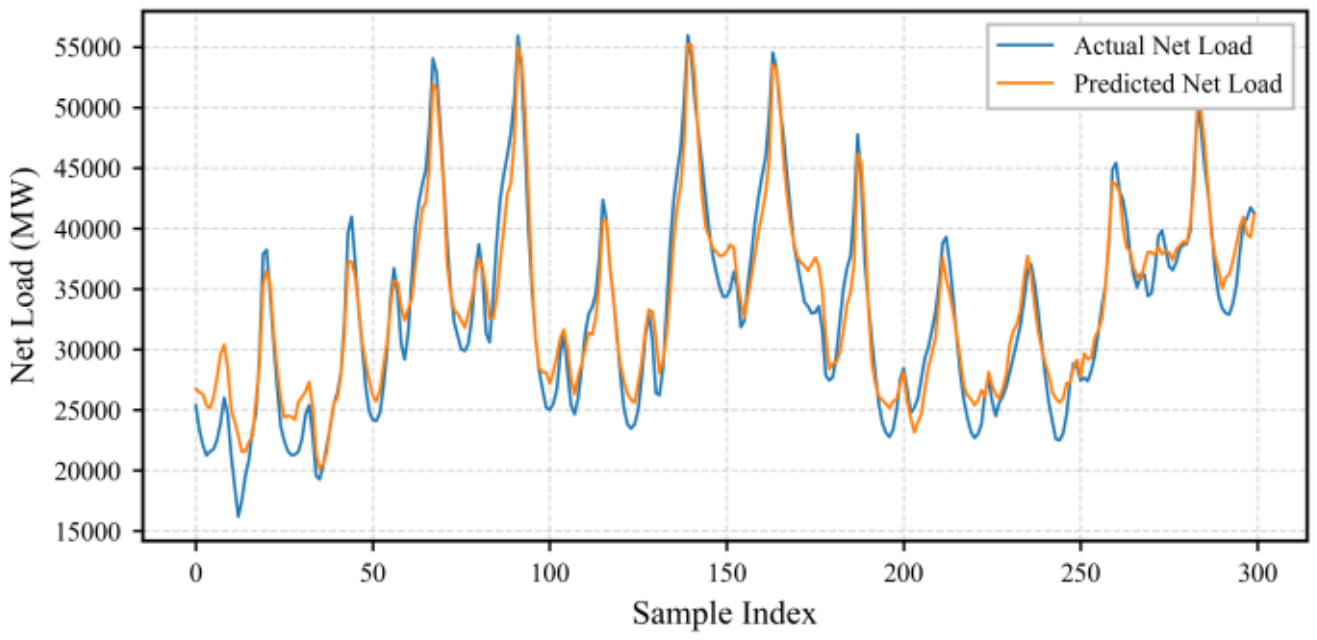

Fig. 7. Comparison of actual and predicted net load for indirect FCNN method.

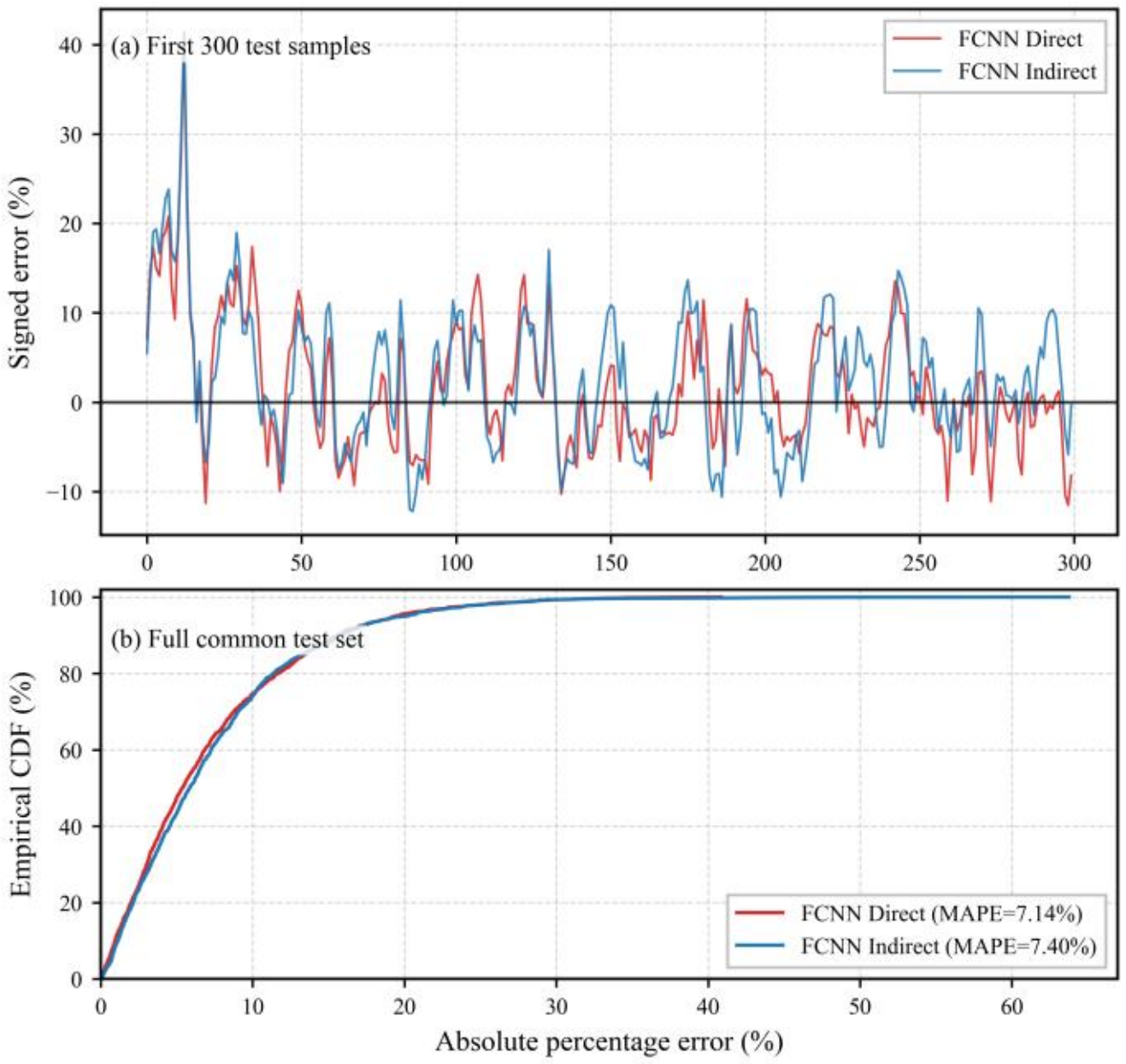

Fig. 8. FCNN forecasting errors: (a) signed percentage errors over the first 300 test samples; and (b) empirical cumulative distributions of absolute percentage errors over the full common test set.

Fig. 8 compares the signed percentage errors of the direct and indirect FCNN methods over the first 300 test samples. Both error sequences fluctuate around zero but exhibit small positive mean biases of 1.73% and 2.67% for the direct and indirect methods, respectively. The indirect FCNN also shows slightly greater dispersion over this subset. Consequently, the corrected results do not indicate persistent underprediction by the indirect method or opposite bias directions between the two FCNN formulations.

## B. *Results of LSTM Model*

After tuning the LSTM model, the direct method produced its final result. As with this model, the parameters that were held constant in other models were also kept constant in this case. Fig. 9 shows the training and validation Huber losses of the direct LSTM model. The model was configured for a maximum of 100 epochs, while early stopping terminated training after 53 epochs and restored the weights from epoch 43, where the minimum validation loss was 0.0038. The training loss decreased from 0.1462 to approximately 0.0076, while the validation loss remained stable at approximately 0.0040 during the final epochs. The lower validation loss is consistent with dropout being active during training but disabled during validation.

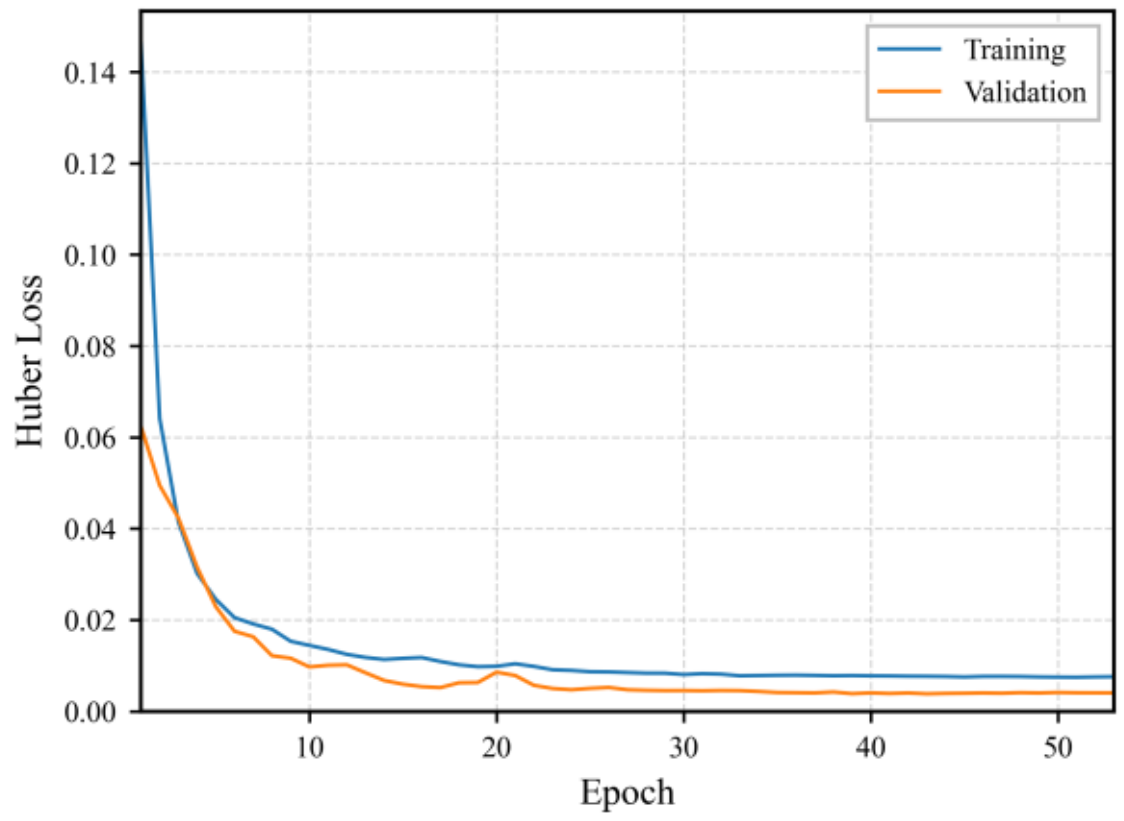

Fig. 9. Training and validation Huber losses of the direct LSTM model.

Table III shows comparable performance between direct LSTM and indirect LSTM models, although the direct method produced modestly lower MAPE and RMSPE values and a higher $R^2$.

Table III. LSTM Performance of both Direct and Indirect methods

| Performance Measure | Direct Method | Indirect Method |
|---|---|---|
| MAPE (%) | 2.73% | 2.46% |
| RMSPE (%) | 3.83% | 3.59% |
| $R^2$ Score | 98.1% | 98.7% |

Fig. 10 shows the actual and predicted net load profiles of the direct LSTM method over the first 300 test samples. The direct LSTM achieved a MAPE of 2.73%, an RMSPE of 3.83%, and an $R^2$ of 98.1%. After running multiple LSTM models for load consumption, solar and wind power produced with the same parameters as the direct LSTM method such as window length, epochs and so on. As shown in Fig. 11, the indirect LSTM predictions also closely follow the actual net load profile. The indirect LSTM achieved a MAPE of 2.46%, an RMSPE of 3.59%, and an $R^2$ of 98.7% as shown in Table III, representing the best aggregate performance among the four evaluated methods.

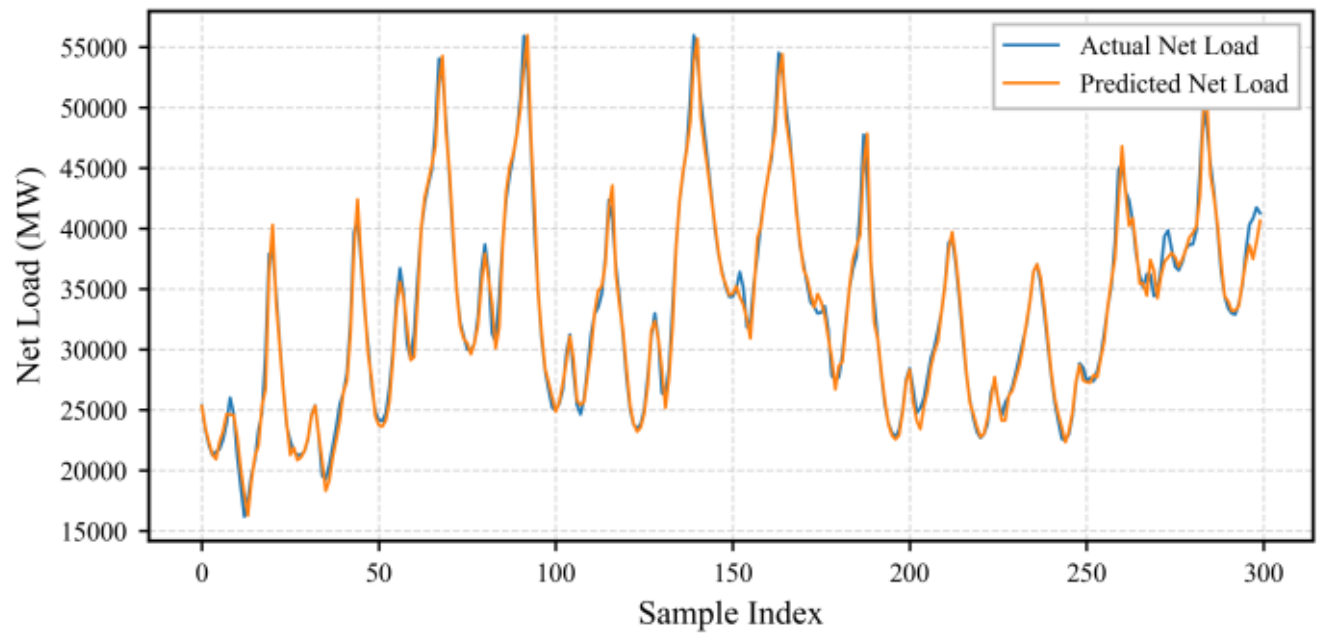

Fig. 10. Comparison of actual and predicted net load for direct LSTM method.

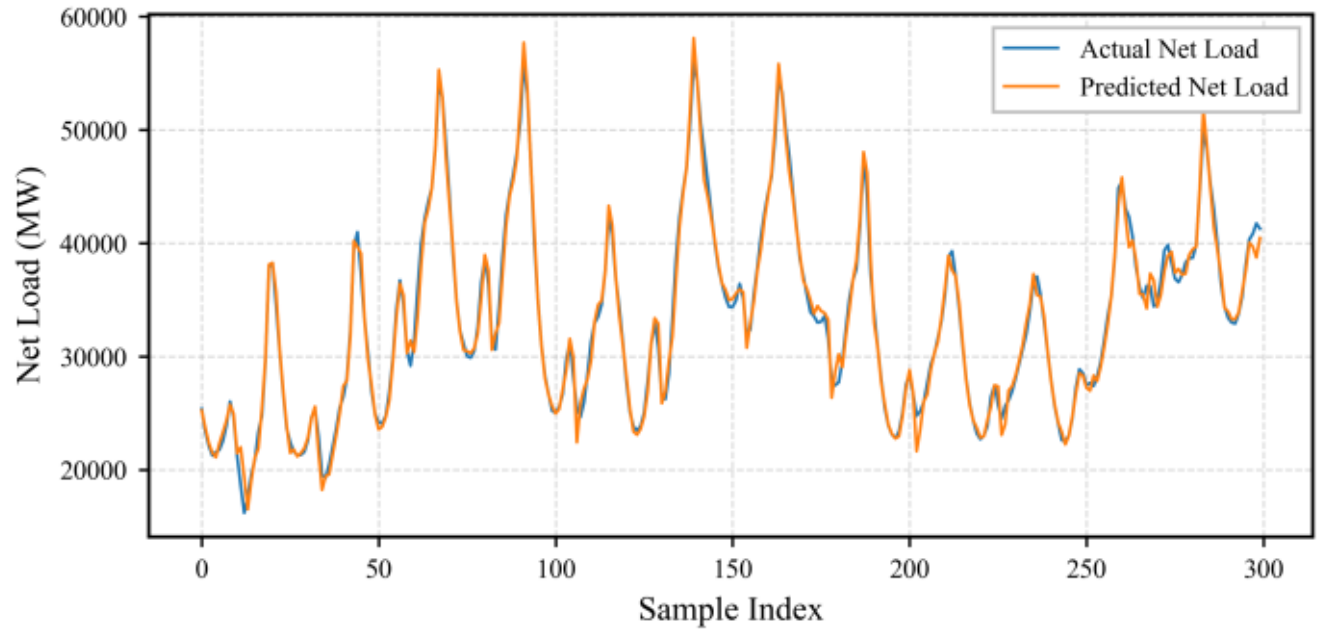

Fig. 11. Comparison of actual and predicted net load for indirect LSTM method.

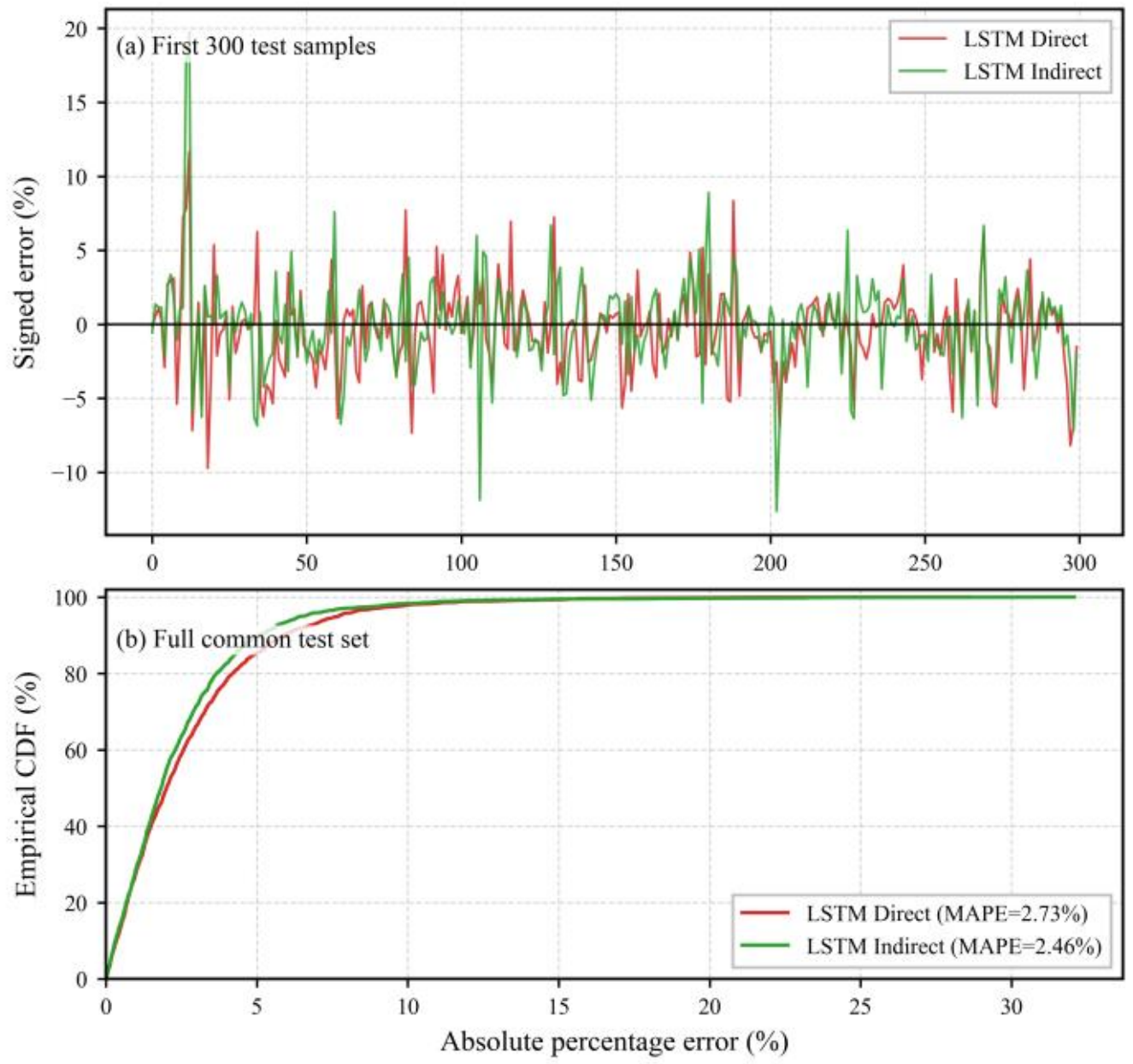

Fig. 12. LSTM forecasting errors: (a) signed percentage errors over the first 300 test samples; and (b) empirical cumulative distributions of absolute percentage errors over the full common test set.

Fig. 12 shows that the signed percentage errors of both LSTM methods are generally centered near zero. Over the first 300 samples, the direct LSTM exhibits a slightly narrower error range, whereas the indirect LSTM contains several larger local deviations. Nevertheless, evaluation over the full common test set shows that the indirect LSTM achieves lower MAPE, RMSPE, median absolute percentage error, and overall APE dispersion. Thus, its advantage is reflected in aggregate performance rather than uniformly smaller errors at every timestamp.

Table IV shows that the indirect LSTM achieved the lowest median absolute percentage error of 1.83% and the lowest APE standard deviation of 2.61%. However, the direct LSTM achieved the lowest maximum absolute percentage error of 28.64%, compared with 32.08% for the indirect LSTM. The indirect FCNN exhibited the largest maximum error of 63.77%. These results indicate that the indirect LSTM provides the best typical and aggregate accuracy, while the direct LSTM has a slightly better worst case percentage error.

Table IV. Statistical Results of Absolute Percentage Errors for the Predictions of the Four Evaluated Methods

| Method Name | Max (%) | Min (%) | Median (%) | Std Dev (%) |
|---|---|---|---|---|
| Direct FCNN | 40.87% | 0 | 5.35% | 6.26% |
| Indirect FCNN | 63.77% | 0 | 5.85% | 6.43% |
| Direct LSTM | 28.64% | 0 | 2.01% | 2.69% |
| Indirect LSTM | 32.08% | 0 | 1.83% | 2.60% |

## IV. Conclusions

This study compares the direct and indirect net load forecasting using FCNN and LSTM models with weather, load, renewable generation, and renewable capacity inputs. The direct FCNN achieved a MAPE of 7.14%, an RMSPE of 9.49%, and an $R^2$ of 92.3%, while the indirect FCNN achieved 7.40%, 9.80%, and 91.1%, respectively. The direct LSTM achieved a MAPE of 2.73%, an RMSPE of 3.83%, and an $R^2$ of 98.1%, while the indirect LSTM achieved 2.46%, 3.59%, and 98.7%, respectively. Overall, the indirect LSTM achieved the best aggregate performance among the four evaluated methods.

These results demonstrate that reliable net load forecasting can be achieved using relatively compact neural network models when renewable generation, weather variables, and renewable capacity features are jointly considered. Although the LSTM-based indirect method achieves the best overall performance among the evaluated cases, this work does not claim that FCNN and LSTM exhaust the full forecasting-model space. Future work will extend the comparison to additional baselines, including GRU, ARIMA, XGBoost, SVM, and Transformer-based models, and will further quantify the contribution of renewable capacity features through ablation analysis.

## Acknowledgment

This work is supported in part by the University of Houston Energy Scholars program.